\documentclass[sigconf]{acmart}
\usepackage{graphicx}
\usepackage{bm}
\usepackage{caption, subcaption}
\usepackage{etoolbox}
\usepackage{enumitem}
\usepackage{multicol}
\usepackage{rotating}
\usepackage{hyperref}
\usepackage{cleveref}
\usepackage{float}
\usepackage{here}

\AtBeginDocument{%
  \providecommand\BibTeX{{%
    \normalfont B\kern-0.5em{\scshape i\kern-0.25em b}\kern-0.8em\TeX}}}

\setcopyright{none}
\renewcommand\footnotetextcopyrightpermission[1]{}

\makeatletter
\renewcommand\@formatdoi[1]{\ignorespaces}
\makeatother

\begin{document}
\title[DeepCOVIDExplainer: Explainable COVID-19 Diagnosis Based on Chest X-ray Images]{DeepCOVIDExplainer: Explainable COVID-19 Diagnosis\texorpdfstring{\\}{} Based on Chest X-ray Images}

\acmConference{\vphantom{BLA}  \begingroup\color{white} }{\vphantom{BLA}}{\vphantom{BLA} \endgroup}

\author{Md. Rezaul Karim}
\affiliation{
    \institution{Fraunhofer FIT, Aachen, Germany}
    \institution{RWTH Aachen University, Germany}
}
\author{Till D{\"o}hmen}
\affiliation{
    \institution{Fraunhofer FIT, Aachen, Germany}
     \institution{RWTH Aachen University, Germany}
}
\author{Dietrich Rebholz-Schuhmann}
\affiliation{
    \institution{German National Library of Medicine, University of Cologne, Germany}
}
\author{Stefan Decker}
\affiliation{
    \institution{Fraunhofer FIT, Aachen, Germany}
    \institution{RWTH Aachen University, Germany}
    }
\author{Michael Cochez}
\affiliation{
    \institution{Dept. of Computer Science, Vrije Universiteit Amsterdam, Netherlands}
}
\author{Oya Beyan}
\affiliation{
    \institution{RWTH Aachen University, Germany}
    \institution{Fraunhofer FIT, Aachen, Germany}
}

\renewcommand{\shortauthors}{Karim, D\"ohmen, Rebholz-Schuhmann, Decker, Cochez, and Beyan}

\begin{abstract}
    Amid the coronavirus disease~(COVID-19) pandemic, humanity experiences a rapid increase in infection numbers across the world. Challenge hospitals are faced with, in the fight against the virus, is the effective screening of incoming patients. One methodology is the assessment of chest radiography~(CXR) images, which usually requires expert radiologists' knowledge. In this paper, we propose an explainable deep neural networks~(DNN)-based method for automatic detection of COVID-19 symptoms from CXR images, which we call \emph{`DeepCOVIDExplainer'}. We used 15,959 CXR images of 15,854 patients, covering normal, pneumonia, and COVID-19 cases. CXR images are first comprehensively preprocessed, before being augmented and classified with a neural ensemble method, followed by highlighting class-discriminating regions using gradient-guided class activation maps~(Grad-CAM++) and layer-wise relevance propagation~(LRP). Further, we provide human-interpretable explanations of the predictions. Evaluation results based on hold-out data show that our approach can identify COVID-19 confidently\footnote{Findings are not externally~(e.g., with radiologists/clinicians) validated yet.} with a positive predictive value~(PPV) of 91.6\%, 92.45\%, and 96.12\%; precision, recall, and F1 score of 94.6\%, 94.3\%, and 94.6\%, respectively for normal, pneumonia, and COVID-19 cases, respectively, making it comparable or improved results over recent approaches. We hope that our findings will be a useful contribution to the fight against COVID-19 and, in more general, towards an increasing acceptance and adoption of AI-assisted applications in the clinical practice.
\end{abstract}


\maketitle
\section{Introduction}
\label{introduction}
The ongoing coronavirus pandemic~(declared a pandemic by the World Health Organization~(WHO) in March 2020) has had a devastating impact on the health and well-being of the global population already~\cite{wang2020covid,gozes2020rapid}. As of June 4, 2020, more than 6.6 million infections of COVID-19 and 388,502 fatalities due to the disease were reported\footnote{\url{https://www.worldometers.info/coronavirus/}}. 
Recent studies show that COVID-19, caused by severe acute respiratory syndrome coronavirus 2~(SARS-CoV-2)~\cite{RT_PCR}, often, but by no means exclusively, affects elderly persons with pre-existing medical conditions~\cite{COVID1,COVID2,COVID3,huang2020clinical,ng2020imaging}.
While hospitals are struggling with scaling up capacities to meet the rising number of patients, it is crucial to make use of the screening methods at hand to identify COVID-19 cases and discriminate them from other conditions~\cite{wang2020covid}. 
The definitive test for COVID-19 is the reverse transcriptase-polymerase chain reaction~(RT-PCR) test~\cite{RT_PCR}, which has to be performed in specialized laboratories and is a labour-intensive process. COVID-19 patients, however, show several unique clinical and para-clinical features, e.g., presenting abnormalities in medical chest imaging with commonly bilateral involvement. The features were shown to be observable on chest X-ray~(CXR) and CT images~\cite{huang2020clinical}, but are only moderately characteristic to the human eye~\cite{ng2020imaging} and not easy to distinguish from pneumonia features. 

AI-based techniques have been utilized in numerous scenarios, including automated diagnoses and treatment in clinical settings. Deep neural networks~(DNNs) have recently been employed for the diagnosis of COVID-19 from medical images, leading to promising results~\cite{huang2020clinical,wang2020covid,ng2020imaging,ozturk2020automated,tabik2020covidgr}. However, many current approaches are ``black box" methods without providing insights into the decisive image features. 
Let's imagine a situation where resources are scarce, e.g., a hospital runs out of confirmatory tests or necessary radiologists are occupied, where AI-assisted tools could potentially help less-specialized general practitioners to triage patients, by highlighting critical chest regions to lead automated diagnosis decision~\cite{wang2020covid}. A fully automated method without the possibility for human verification would, however, at the current state-of-the-art, be unconscionable and potentially dangerous in a practical setting. 
As a first step towards an AI-based clinical assistance tool for COVID-19 diagnosis, we propose \emph{`DeepCOVIDExplainer'}, a novel diagnosis approach based on neural ensemble method. The pipeline of \emph{`DeepCOVIDExplainer'} starts with histogram equalization enhancement, filtering, and unsharp masking of CXR images, followed by the training of DenseNet, ResNet, and VGGNet architectures in a transfer learning~(TL) setting, creating respective model snapshots. Those are incorporated into an ensemble, using Softmax class posterior averaging~(SCPA) and prediction maximization~(PM) for the best performing models. 

Finally, class-discriminating attention maps are generated using gradient-guided class activation maps~(Grad-CAM++) and layer-wise relevance propagation~(LRP) to provide explanations of the predictions and to identify the critical regions on patients chest. 
We hope that \emph{`DeepCOVIDExplainer'} will be a useful contribution towards the development and adoption of AI-assisted diagnosis applications in general, and for COVID-19 in particular. 
To allow for the reproduction of results and derivative works, we will make the source code, documentation and links to used data publicly available. 
The rest of the paper is structured as follows: \Cref{sec:rw} outlines related works and points out potential limitations. \Cref{sec:mm} describes our proposed approach, before demonstrating experiment results in \cref{sec:er}. \Cref{sec:co} summarizes the work and provides some outlook before concluding the paper.

\section{Related Work}
\label{sec:rw}
Bullock et al.~\cite{bullock2020mapping} provide a comprehensive overview of recent application areas of AI against COVID-19, mentioning medical imaging for diagnoses first, which emphasizes the prevalence of the topic. Although PCR tests offer many advantages over CXR and CT~\cite{COVID3}, shipping the sample of patients is necessary, whereas X-ray or CT machines are readily available even in rather remote areas. In a recent study by K. Lee et al.~\cite{COVID1}, CXR and CT images from nine COVID-19 infected patients were analyzed by two radiologists to assess the correspondence of abnormal findings on X-rays with those on CT images. The proportion of patients with abnormal initial radiographic findings was 78.3\% to 82.4\% for SARS and 83.6\% for MERS, while being only 33\% for COVID-19 cases~\cite{COVID1}. 
Chest CT images, in contrast, showed double lung involvement in eight out of nine patients. In other words, X-ray may not be the best imaging method for detecting COVID-19, judging by the small cohort of nine patients~\cite{COVID1}. Another study by Yicheng Fang et al.~\cite{COVID2}, however, supports those findings and argues in favour of the effectiveness of CT over X-ray. CT should hence cautiously be considered as the primary imaging source for COVID-19 detection in epidemic areas~\cite{COVID3}. Nevertheless, the limited patient cohort size leaves room for statistical variability and, in contrast to those findings, a few other studies have reported rather promising results for the diagnosis based on CXR imaging~\cite{wang2020covid,narin2020automatic,ghoshal2020estimating}.

Narin et al. \cite{narin2020automatic} evaluated different convolutional neural networks~(CNN) for the diagnosis of COVID-19 and achieved an accuracy of 98\% using a pre-trained ResNet50 model. However, the classification problem is overly simplified by only discriminating between healthy and COVID-19 patients, disregarding the difficulty of distinguishing regular pneumonia conditions from COVID-19 conditions.  
Wang et al. \cite{wang2020covid} proposed COVID-Net to detect distinctive abnormalities in CXR images among samples of patients with non-COVID-19 viral infections, bacterial infections, and healthy patients. 
On a test sample of 92 positive COVID-19 cases among approx. 300 other cases, COVID-Net reaches an overall accuracy of 92.6\% with 97.0\%, 90.0 \%, and 87.1 \% sensitivity for normal, Non-COVID-19, and COVID-19 cases, respectively. On the other hand, COVID-Net achieves PPV of 90.5\%, 91.3\%, and 98.9\% for normal, Non-COVID-19, and COVID-19 cases, respectively. Still, the small sample size does not enable generalizable statements about the reliability of the method. Highlighted regions using `GSInquire' are also not well-localized to critical areas. Overall, training on imbalance data, lack of thorough image preprocessing, and poor decision visualization have hindered this approach. 

In another study by Ozturk et al.~\cite{ozturk2020automated}, a deep learning model called DarkCovidNet is proposed for the automatic diagnosis of COVID-19 based on CXR images. Trained only 125 CXR images DarkCovidNet model to provide COVID-19 diagnosis in two ways: i) binary classification showing COVID-19 vs no-findings and multiclass classification showing COVID-19 vs no-findings vs pneumonia, giving an accuracy of 98.08\% and 87.02\% for binary and multiclass classification settings, respectively. Although the end-to-end COVID-19 diagnostic pipeline is very comprehensive and backed by `you only look once'~(YOLO) real-time object detection system, it has two potential limitations, including a severely low number of image samples and imprecise localization on the chest region. 
Biraja G. et al. \cite{ghoshal2020estimating} employed uncertainty estimation and interpretability based on Bayesian approach to CXR-based COVID-19 diagnosis, which shows interesting results. The results may, however, be impaired by a lack of appropriate image preprocessing and the resulting attention maps show somewhat imprecise areas of interest. In a very recent approach, Tabik et al.~\cite{tabik2020covidgr} curated a richer dataset called COVIDGR-1.0 containing 377 positive and 377 negative PA (Postero Anterior) CXR views. They subsequently, proposed COVID Smart Data based Network~(aka. COVID-SDNet). As claimed, their approach reaches an accuracy of 97.37\%$±$1.86\%,88.14\%$±$2.02\%, 66.5\%$±$8.04\% in severe, moderate and mild COVID severity levels. 

Although these results look promising when compared to expert radiologist sensitivity of 69\%~\cite{tabik2020covidgr}, in most of the cases, the reliability can be questioned for three main reasons: i) the datasets used are severely biased due to a deficient number of COVID-19 cases~\cite{tabik2020covidgr}. Secondly, some results are not statistically reliable and lack of decision biases given the fact that they were mostly made based on a single model's outcome. Thirdly, less accurate localization and visualization of critical chest regions.
Therefore, to overcome these shortcomings of state-of-the-art approaches, our approach first enriches existing datasets with more COVID-19 samples, followed by a comprehensive preprocessing pipeline for CXR images and data augmentation. The COVID-19 diagnosis of \emph{`DeepCOVIDExplainer'} is based on snapshot neural ensemble method with a focus on fairness, algorithmic transparency, and explainability, with the following assumptions:

\begin{itemize}
    \item By maximum~(or average) voting from a panel of independent radiologists~(i.e., ensemble), we get the final prediction fair and trustworthy than a single radiologist.
    \item By localizing class-discriminating regions with Grad-CAM++ and LRP, we not only can mitigate the opaqueness of the black-box model by providing more human-interpretable explanations of the predictions but also identify the critical regions on patients chest.
\end{itemize}

\section{Methods}
\label{sec:mm}
In this section, we discuss our approach in detail, covering network construction and training, followed by the neural ensemble and decision visualizations. 

\subsection{Preprocessing}
Depending on the device type, radiographs almost always have dark edges on the left and right side of the image. Hence, we would argue that preprocessing is necessary to make sure the model not only learns to check if the edges contain black pixels or not but also to improve its generalization. 
We perform contrast enhancement, edge enhancement, and noise elimination on entire CXR images by employing histogram equalization~(HGE), Perona-Malik filter~(PMF), and unsharp masking edge enhancement. Since images with distinctly darker or brighter regions impact the  classification~\cite{86}, we perform the global contrast enhancement of CXR images using HGE. By merging gray-levels with low frequencies into one, stretching high frequent intensities over high range of gray levels, HGE achieves close to equally distributed intensities~\cite{90}, where the probability density function $p(X_{k})$ of an image $X$ is defined as~\cite{90}:

\begin{equation}
    p(X_{k})=\frac{n_{k}}{N},
    \label{eq3.1}
\end{equation}

\noindent where $k$ is the grey-level ID of an input image $X$ varying from 0 to $L$, $n_{k}$ is the frequency of a grey level $X_{k}$ appearing in $X$, and $N$ is the number of image samples. A plot of $n_{k}$ vs. $X_{k}$ is specified as the histogram of $X$, while the equalization transform function $f(X_{k})$ is tightly related to cumulative density function~\cite{90}:

\begin{align}
    f(X_{k})=X_{0}+(X_{L}-X_{0})c(X_{k}) \\
    c(X_{k})=\sum_{j=0}^{k}p(X_{j}).
\end{align}

\noindent Output of HGE, $Y={Y(i,j)}$ is synthesized as follows~\cite{90}:

\begin{equation}
    Y=f(X)=\left\{f(X(i,j))|\forall X(i,j) \in X\right\}.
\end{equation}  

\vspace{2mm}
Image filters `edge enhances' and `sharpen' are adopted with the convolution matrices as kernel $g(.)$. PMF is used to preserve the edges and detailed structures along with noise reduction as long as the fitting diffusion coefficient $c(.)$ and gradient threshold $K$ are separate~\cite{95}. As a non-linear anisotropic diffusion model, PMF smoothens noisy images $\theta (x,y)$ w.r.t. partial derivative as~\cite{95}:

\begin{equation}
    \frac{\partial u}{\partial t}= div(c(|\nabla u(x,y,t)|)\nabla u(x,y,t)),
\end{equation}

\vspace{2mm}
\noindent  where $u(x,y,0)$ is the original image, $\theta (x,y)$, $u(x,y,t)$ is a filtered image after $t$ iteration diffusion; $div$ and $\nabla$ are divergence and gradient operators w.r.t spatial variables $x$ and $y$, where the diffusion coefficient $c(.)$ is computed as~\cite{96}:

\begin{align}
    c_{1}(|\nabla I|)=exp \left(- \left(\frac{|\nabla I|}{K} \right)^{2} \right) \\
    c_{2}(|\nabla I|)=\frac{1}{1+\left(\frac{|\nabla I|}{K} \right)^{2}}.
\end{align}

To decide if the local gradient magnitudes is robust for edge preservation, diffusion coefficient function $c(.)$ is then computed as follows~\cite{96}: 

\begin{equation}
    c_{3}(|\nabla I|)=\left\{
    \begin{array}{lr}
    \frac{1}{2}\left[1-\left(\frac{|\nabla I|}{K\sqrt{2}}\right)^{2}\right]^{2}, &|\nabla I|\leq K\sqrt{2}\\
    0,& |\nabla I|> K\sqrt{2}
    \end{array}
    \right.
    \label{eq3.9}
\end{equation}

\noindent where $c_{3}$ is the Tukey's biweight function. Since the boundary between noise and edge is minimal, $c_{3}$ is applied as the fitting diffusion coefficient~\cite{95}. Further, we attempt to remove textual artefacts from CXR images, e.g., a large number of images annotate right and left sides of chest with a white `R' and `L' characters. To do so, we threshold the images first to remove very bright pixels, and the missing regions were in-painted. 
In all other scenarios, image standardization and normalization are performed. For image standardization, the mean pixel value is subtracted from each pixel and divided by the standard deviation of all pixel values. The mean and standard deviation is calculated on the whole datasets and adopted for training, validation and test sets. For image normalization, pixel values are rescaled to a [0,1] by using a pixel-wise multiplication factor of 1/255, giving a collection of grey-scale images. Further, CXR images are resized $224 \times 224 \times 3$ before starting the training. 

\begin{figure*}
	\centering
	\includegraphics[width=0.9\linewidth,height=65mm]{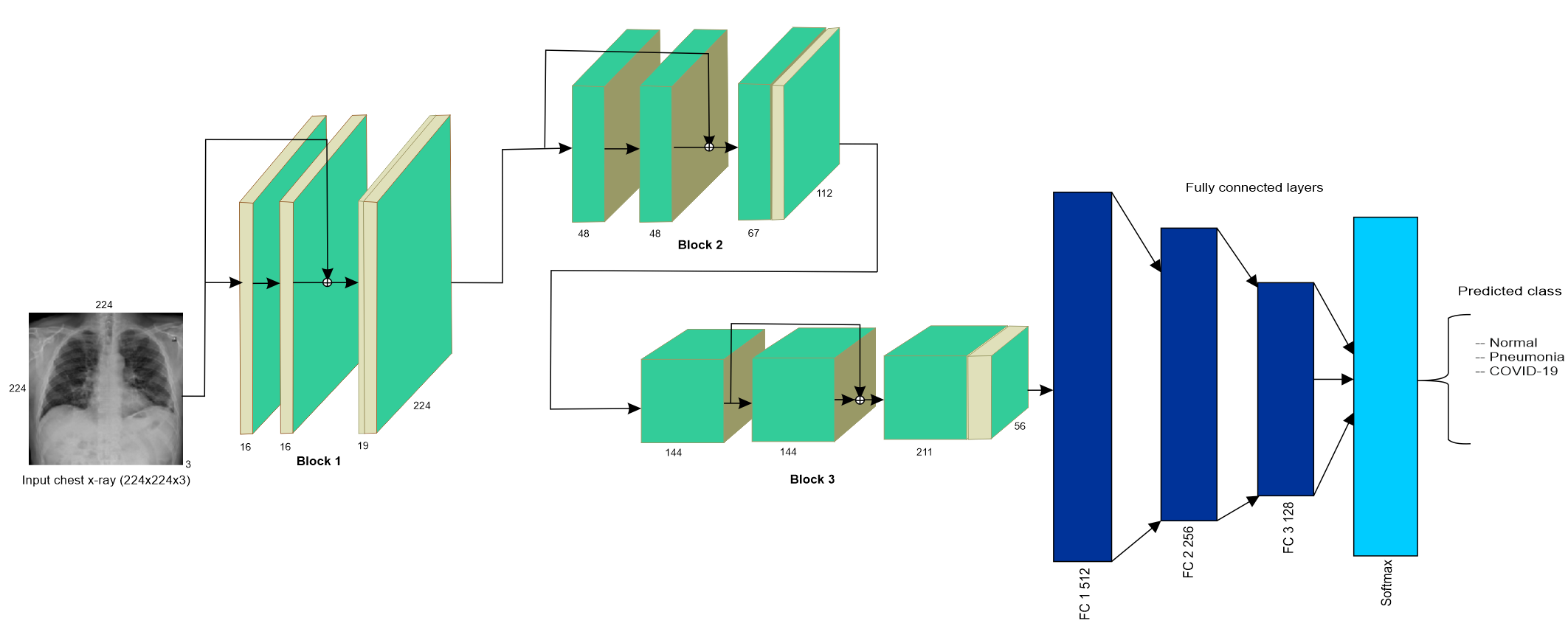}
	\caption{The classification with ResNet-based networks} 
	\label{fig:resnet}
\end{figure*}

\subsection{Network construction and training}
\begin{figure*}
    \centering
    \includegraphics[width=0.8\textwidth,height=60mm]{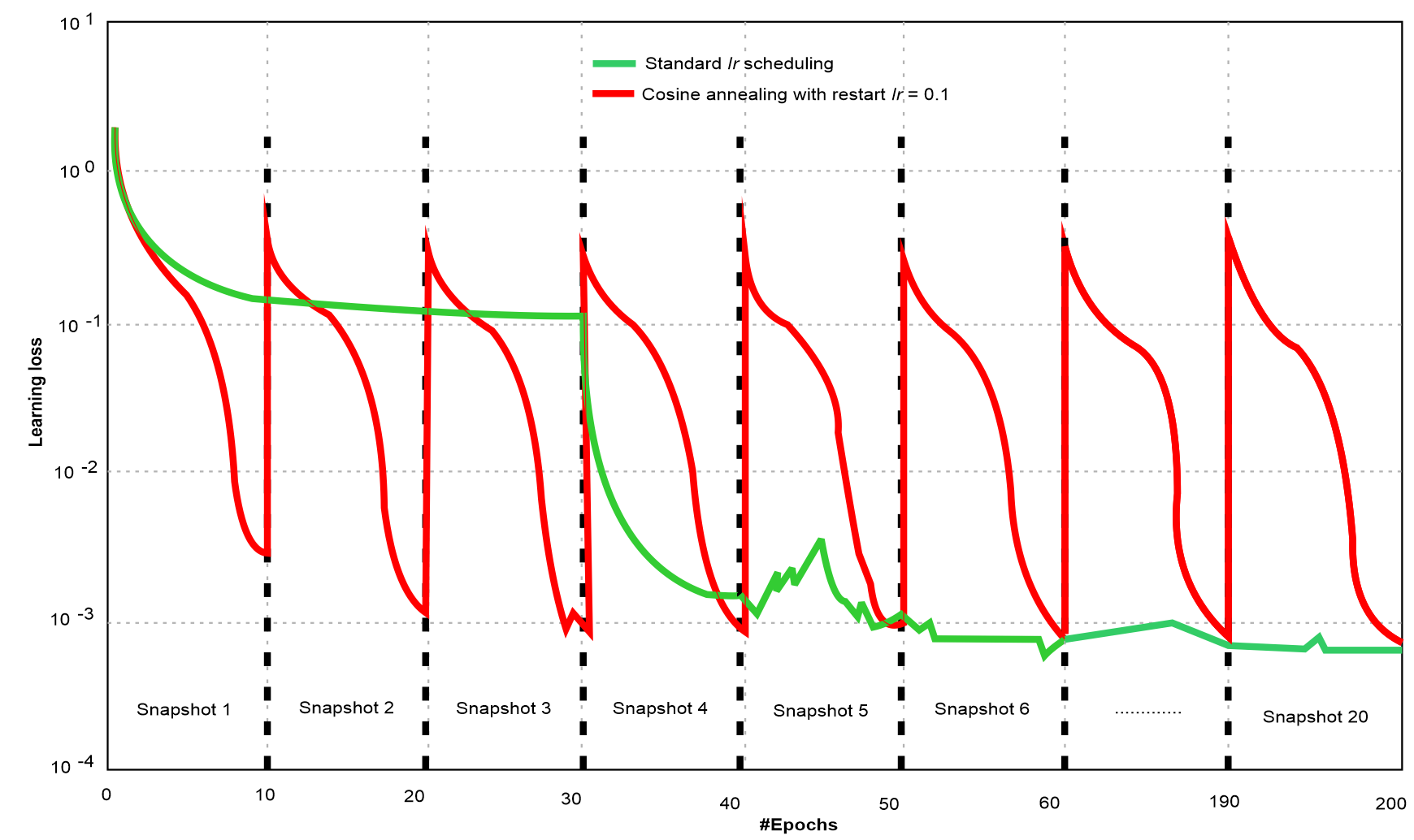}
    \caption{Training loss of VGG-19 network with standard learning rate~(green) and cosine annealing cycles~(red), the intermediate models, denoted by the dotted lines form an ensemble at the end of training}
    \label{fig:cac}
\end{figure*}

We trained VGG, ResNet, and DenseNet architectures and created several snapshots during a single training run with cyclic cosine annealing~(CAC)~(see \cref{fig:cac})\cite{loshchilov2016sgdr}, followed by combining their predictions to an ensemble prediction~\cite{huang2017snapshot,karim2019snapshot}. We pick VGG-16 and VGG-19 due to their general suitability for image classification. Based on the dense evaluation concept~\cite{108}, VGG variants convert the last three fully connected layers~(FCLs) to 2D convolution operations to reduce the number of hyperparameters. We keep last 2 layers fixed to adopt a 1$\times$1 kernel, leaving the final one equipped with a Softmax activation.
However, owing to computational complexity of VGG-16 due to consecutive FCLs, the revised VGG-19 is trained with a reduced number of hidden nodes in first 2 FCLs.

Next, we pick ResNet-18~\cite{107}) and ResNet-34~\cite{106}) architectures. ResNets are lightweight stack-based CNNs, with their simplicity arising from small filter sizes~(i.e., 3$\times$3)~\cite{108}, where apart from common building blocks, two bottlenecks are present in the form of channel reduction. 
A series of convolution operators without pooling is placed in between, forming a stack. The first conv layer of each stack in ResNets~(except for the first stack) are down-sampled at stride 2, which provokes the channel difference between identity and residual mappings. 
A series of convolution operators without pooling is then placed in between and recognized as a stack, as shown in \cref{fig:resnet}.
However, w.r.t regularisation, a 7$\times$7 conv filter is decomposed into a stack of three 3$\times$3 filters with non-linearity injected in between~\cite{108}. DenseNet-161 and DenseNet-201 architectures are picked that concatenate additional inputs from preceding layers, while ResNets merge feature-maps through summation. It not only strengthens feature propagation and moderates information loss but also increases feature reusing capability by reducing numbers of parameters~\cite{109}.  

To avoid possible overfitting, $L_2$ weight regularization, dropout, and data augmentation~(by rotating CXR images by up to 15$^\circ$) were employed. We do not initialize networks weights with any pretrained~(e.g., ImageNet) models. ImageNet contains photos of general objects, which would activate the internal representation of the network's hidden layers with geometrical forms, colourful patterns, or irrelevant shapes that are usually not present CXR images. 
We set the number of epochs~(NE), maximum learning rate~(LR), number of cycles, and current epoch number, where initial LR and NE are two hyperparameters. CAC starts with a large LR and rapidly decreases to a minimum value before it dramatically increases to the following LR for that epoch~\cite{huang2017snapshot}. During each model training, CAC changes the LR aggressively but systematically over epochs to produce different network weights~\cite{huang2017snapshot}: 

\begin{equation}
    \label{eq:lr-cosine}
    \alpha(t)=\frac{\alpha_{0}}{2}\left(\cos \left(\frac{\pi \bmod (t-1,\lceil T / C\rceil)}{\lceil T / C\rceil}\right)+1\right),
\end{equation}

where $\alpha(t)$ is the LR at epoch $t$, $\alpha_0$ is the maximum LR, $T$ is the total epoch, $C$ is the number of cycles and $mod$ is the modulo operation. After training a network for $C$ cycles, best weights at the bottom of each cycle are saved as a model snapshot~($m$), giving $M$ model snapshots, where $m \leq M$. 

\subsection{Model ensemble}
\noindent When a single practitioner makes a COVID-19 diagnosis, the chance of a false diagnosis is high. In case of doubt, a radiologist should, therefore, ask for a second or third option of other experts. Analog to this principle, we employ the principle of model ensembles, which combine the `expertise' of different predictions algorithms into a consolidated prediction and hereby reducing the generalization error~\cite{huang2017snapshot}. Research has shown that a neural ensemble method by combining several deep architectures is more effective than structures solely based on a single model~\cite{huang2017snapshot,karim2019snapshot}. 
Inspired by~\cite{7}, we apply both SCPA and PM of best-performing snapshot models, ensemble their predictions, and propagate them through the Softmax layer, where the class probability of the ground truth $j$ for a given image $x$ is inferred as follows~\cite{7}:

\begin{equation}
    P(y=j | \mathbf{x})=\frac{\exp \left[\sum_{m=1}^{M} \hat{P}_{m}(y=j | \mathbf{x})\right]}{\exp \left[\sum_{k=1}^{K} \sum_{m=1}^{M} \hat{P}_{m}(y=k | \mathbf{x})\right]},
\end{equation}

\noindent where $m$ is the last snapshot model, $M$ is the number of models, $K$ is the number classes, and $\hat{P}_{m}(y=j | \mathbf{x})$ is the probability distribution. 

\begin{figure*}
	\centering
	\includegraphics[width=0.95\textwidth,height=60mm]{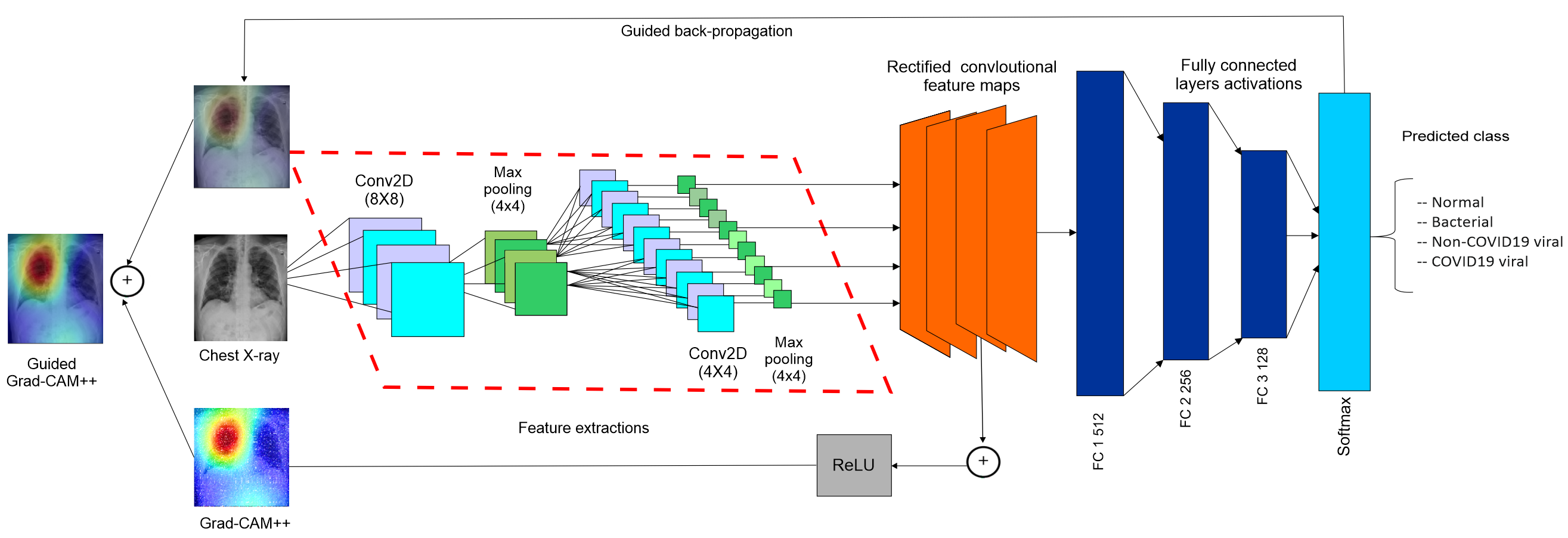}
    \caption{Classification and decision visualization with CNN-based approach}	
	\label{fig:viz}
\end{figure*}

\subsection{Decision visualizations}
To improve the COVID-19 detection transparency, class-discriminating regions on the subjects chest are generated by employing Grad-CAM~\cite{114}, Grad-CAM++~\cite{chattopadhay2018grad}, and LRP~\cite{LRP2}. The idea is to explain where the model provides more attention for the classification. CAM computes the number of weights of each feature map~(FM) based on the final conv layer to calculate the contribution to prediction $y_c$ at location $(i,j)$, where the goal is to obtain $L_{ij}^{c}$ that satisfies $y^{c}=\sum_{i, j} L_{ij}^{c}$. The last FM $A_{ijk}$ and the prediction $y_c$ are represented in a linear relationship in which linear layers consist of global average pooling~(GAP) and FCLs: i) GAP outputs $F_{k}=\sum_{i,j} A^k_{ij}$, ii) FCL that holds weight $w_{k}^{c}$, generates the following output~\cite{kim2020extending}: 
 
 \begin{align}
     y^{c}=\sum_{k} w_{k}^{c} F_{k}=\sum_{k} w_{k}^{c} \sum_{i,j} A^k_{ij}=\sum_{i, j} \sum_{k} w_{k}^{c} A^k_{ij},
 \end{align}
 
where $A^{k}$ represents the visualization of the $k^{th}$ feature map, $L_{i j}^{c}=\sum_{k} w_{k}^{c} A^k_{ij}$~\cite{kim2020extending}. Due to the vanishing of non-linearity of classifiers, CAM is an unsuitable method. Hence, we employ Grad-CAM to globally average the gradients of FM as weights instead of pooling. 
While heat maps~(HM) are plotted, class-specific weights are collected from the final conv layer through globally averaged gradients~(GAG) of FM instead of pooling~\cite{chattopadhay2018grad}: 

\begin{equation}
    \alpha_k^c=\frac{1}{Z}\sum_{i}\sum_{j}\frac{\partial y^c}{\partial A_{ij}^k},
    \label{eq:alpha}
\end{equation}

where $Z$ is the number of pixels in an FM, $c$ is the gradient of the class, and $A_{ij}^k$ is the value of $k^{th}$ FM. Having gathered relative weights, the coarse saliency map~(SM), $L^c$ is computed as the weighted sum of $\alpha_k^c*A_{ij}^k$ of the ReLU activation.
It introduces linear combination to the FM as only the features with a positive influence on the respective class are of interest~\cite{chattopadhay2018grad} and the negative pixels that belong to other categories in the image are discarded~\cite{114}:

\vspace{-2mm}
\begin{equation}
    L^c=\operatorname{ReLU}(\sum_{i}\alpha_k^cA^k),
    \label{3.11}
\end{equation}
\vspace{-2mm}

Grad-CAM++~(see \cref{fig:viz}) replaces the GAG with a weighted average of the pixel-wise gradients as the weights of pixels contribute to the final prediction w.r.t the following iterators over the same activation map $A^k$, $(i,j)$ and $(a,b)$.

\begin{align}
    w_k^c=\sum_{i}\sum_{j}\alpha_{ij}^{kc}\cdot \operatorname{ReLU}(\frac{\partial y^c}{\partial A_{ij}^k}) \\ 
    y^c=\sum_{k}w_k^c\cdot \sum_{i}\sum_{j}A_{ij}^k \\
    \alpha_{ij}^{kc}=\frac{\frac{\partial^2y^c}{(\partial A_{ij}^k)^2}}{2\frac{\partial^2y^c}{(\partial A_{ij}^k)^2}+\sum_{a}\sum_{b}A_{ab}^k{\frac{\partial^3y^c}{\{(\partial A_{ij}^k)^3\}}}}.
    \label{eq:w}
\end{align}

Even though CXR images rarely contain multiple targets, revealing particular image parts that contributed to the prediction, rather than the entire chest area is still helpful. CAM variants back-propagate the gradients all the way up to the inputs, are essentially propagated only till the final conv layer. Besides, CAM methods are limited to specific architectures, where an average-pooling layer connects conv layers with an FCL. LRP is another robust technique of propagating relevance scores~(RSs) and, in contrast to CAM, redistributes proportionally to the activation of previous layers. LRP assumes that the class likelihood can be traced backwards through a network to the individual layer-wise nodes of the input~\cite{LRP2}. From a network of $L$ layers, $1,2,...,N$ nodes in layer $l$, $1,2,..,M$ nodes in layer $l+1$, the RS, $R_{n}^{(l)}$ at node $n$ in layer $l$ is recursively defined~\cite{LRP2}:   

\begin{align}
    R_{n}^{(l)}=\sum_{m} \frac{a_{n}^{(l)} w_{n, m}^{+(l)}}{\sum_{n^{\prime}} a_{n^{\prime}}^{(l)} w_{n^{\prime}, m}^{+(l)}} R_{m}^{(l+1)}.
\end{align}
\vspace{2mm}

\noindent Node-level RS for negative values is calculated with ReLU activation function as follows~\cite{LRP2}:

\begin{align}
    R_{n}^{(l)}=\sum_{m} \frac{x_{n}^{(l)} w_{n, m}^{(l)}-b_{n}^{(l)} w_{n, m}^{+(l)}-h_{n}^{(l)} w_{n, m}^{-(l)}}{\sum_{n^{\prime}} x_{n^{\prime}}^{(l)} w_{n^{\prime}, m}^{(l)}-b_{n^{\prime}}^{(l)} w_{n^{\prime}, m}^{+(l)}-h_{n^{\prime}}^{(l)} w_{n^{\prime}, m}^{-(l+1)}}.
\end{align}
\vspace{2mm}

\noindent Then the output layer RS is calculated before being back-propagated as follows~\cite{LRP2}:

\begin{align}
    R_{n}^{(L)}=\left\{\begin{array}{ll}
    {z_{t}^{(L)}} & {n=t} \\
    {0} & {\text { otherwise }}
    \end{array}\right.
\end{align}
\vspace{2mm}

First, an image $x$ is classified in a forward pass, where LRP identifies important pixels. The backward pass is a conservative relevance~(i.e., $R_{t}^{(L)}$) redistribution procedure with back-propagation using deep Taylor decomposition~\cite{DTD}, to generate a relevance map $R_{lrp}$, for which the nodes contributing most to the higher-layer, also receive most relevance. Finally, heat maps for all the test samples are generated based on the trained models, indicating the relevance for the classification decision. 

\section{Experiment Results}
\label{sec:er}
In this section, we discuss our evaluation results both quantitative and qualitatively, showing a comparative analysis. 

\subsection{Experiment setup}
Experiments were carried out on a machine having an Intel(R) Xeon(R) E5-2640, 256 GB of RAM, and Ubuntu 16.04 OS. All the programs\footnote{\url{https://github.com/rezacsedu/DeepCOVIDExplainer}} were written in Python, where the software stack consists of scikit-learn and Keras with the TensorFlow backend. The LRP-based visualization and relevance calculation are generated using the iNNvestigate toolbox\footnote{\url{https://github.com/albermax/innvestigate}}.
Networks were trained on an Nvidia Titan Xp GPU with CUDA, and cuDNN enabled to make the overall pipeline faster. 
When we create snapshots, we set the number of epochs to 200, maximum LR to 1.0, and the number of cycles to 20, giving 20 snapshots for each model. For 6 architectures, we get 120 snapshot models in total, on which we construct the ensemble. The best snapshot model is used for the decision visualizations, which we choose using WeightWatcher. 

To tackle the class imbalance issue, we apply class weighting to penalize a model when it misclassifies a positive sample. 
Although accuracy is an intuitive evaluation criterion for many bio-imaging problems, e.g., osteoarthritis severity prediction \cite{58}, those evaluation criteria are most suitable for balanced class scenarios. Keeping in mind the imbalanced scenario with widely different class distributions between classes, we report precision, recall, F1, and positive predictive value~(PPV) produced through random search and 5-fold cross-validation tests, i.e., for each hyperparameter group of the specific network structure, 5 repeated experiments are conducted.

\subsection{Datasets}
\label{sec:ds}
We consider 2 different versions of the datasets: first, we used the \emph{`COVIDx v1.0'} dataset by Wang et al.~\cite{wang2020covid} used to train and evaluate the COVID-Net, comprised of a total of 13,975 CXR images across 13,870 patient cases\footnote{As of June 6, 2020}. COVIDx is mainly based on RSNA Pneumonia Detection Challenge\footnote{\url{https://www.kaggle.com/c/rsna-pneumonia-detection-challenge}}, ActualMed COVID-19 Chest X-ray Dataset Initiative\footnote{\url{https://github.com/agchung/Figure1-COVID-chestxray-dataset}}, COVID-19 radiography database\footnote{\url{https://www.kaggle.com/tawsifurrahman/covid19-radiography-database}}, giving 219 COVID-19 positive images, 1,341 normal images, and 1,345 viral pneuomonia images. This gives 358 CXR images from 266 COVID-19 patient cases and total of 8,066 patient cases who have no pneumonia~(i.e., normal) and 5,538 patient cases who have non-COVID19 pneumonia. 
The updated dataset, which we refer to \emph{`COVIDx v2.0'} is categorized as normal~(i.e., no-findings), pneumonia, and COVID-19 viral are enriched with CXR images of adult subjects of COVID-19, pneumonia, and normal examples, leaving 15,959 CXR images across 15,854 patients: 

\vspace{1mm}
\begin{itemize}
    \item {COVID chest X-ray-dataset}: Joseph P.C. et al.~\cite{cohen2020covid}\footnote{\url{https://github.com/ieee8023/covid-chestxray-dataset}}: 660 PA~(i.e., frontal view) CXR images. 
    \item {COVID-19 patients lungs X-ray images}\footnote{\url{https://www.kaggle.com/nabeelsajid917/covid-19-x-ray-10000-image}}: 70 COVID-19 and 70 normal CXR images. 
    \item {Chest X-ray images} by Ozturk et al.~\cite{ozturk2020automated} \footnote{\url{https://github.com/muhammedtalo/COVID-19}}: 125 COVID-19, 500 normal, and 500 pneumonia CXR images. 
\end{itemize}
\vspace{1mm}

\subsection{Performance of individual model}
Overall results are summarized in \cref{Table:all_models}: VGG-19 and DenseNet-161 performed best on both balanced and imbalanced datasets, while VGG-16 turns out to be the lowest performer. In direct comparison, diagnosis of VGG-19 yields much better results than VGG-16, which can be explained by the fact that a classifier with more formations requires more fitting of FMs, which again depends on conv layers. The architecture modification of VGG-19 by setting 2 conv layers and the filter size of 16, visibly enhances the performance. ResNet-18 performed better, although it's larger counterpart ResNet-34 shows very unexpected low performance. Evidently, due to structured residual blocks, the accumulation of layers could not promote FMs extracted from the CXR images.

\begin{table*}
    \centering
	\caption{Classification results of each model on balanced and imbalanced datasets} 
	\label{Table:all_models}
	\begin{tabular}{p{2.9cm}p{1.6cm}p{1.4cm}p{1.0cm}p{0.3cm}p{1.5cm}p{1.4cm}p{1.0cm}}
	 & \multicolumn{3}{c}{\bfseries{Balanced dataset}} && \multicolumn{3}{c}{\bfseries{Imbalanced dataset}} \\
		\cmidrule{2-4}\cmidrule{6-8}   
        \textbf{Network}& \textbf{Precision} & \textbf{Recall}& \textbf{F1} && \textbf{Precision} & \textbf{Recall}& \textbf{F1}\\
		\hline
		VGG-16 & 0.783 & 0.771 & 0.761 && 0.734 & 0.753 & 0.737\\
		ResNet-34 & 0.884 & 0.856 & 0.861 && \textbf{0.852} & \textbf{0.871} & \textbf{0.851} \\
		DenseNet-201  & 0.916 & 0.905 & 0.905 && 0.805 & 0.773 & 0.826\\		
		ResNet-18 & 0.924 & 0.925 & 0.921 && 0.873 & 0.847 & 0.852\\
		VGG-19 & \textbf{0.943} & \textbf{0.935} & \textbf{0.925} && 0.862 & 0.848 & 0.845\\
		DenseNet-161  & \textbf{0.952} & \textbf{0.945} & \textbf{0.948} && \textbf{0.893} & \textbf{0.874} & \textbf{0.883}\\
		\hline
	\end{tabular}
\end{table*}

\begin{figure}
   \centering
    \frame{
    \includegraphics[width=0.43\textwidth,height=55mm]{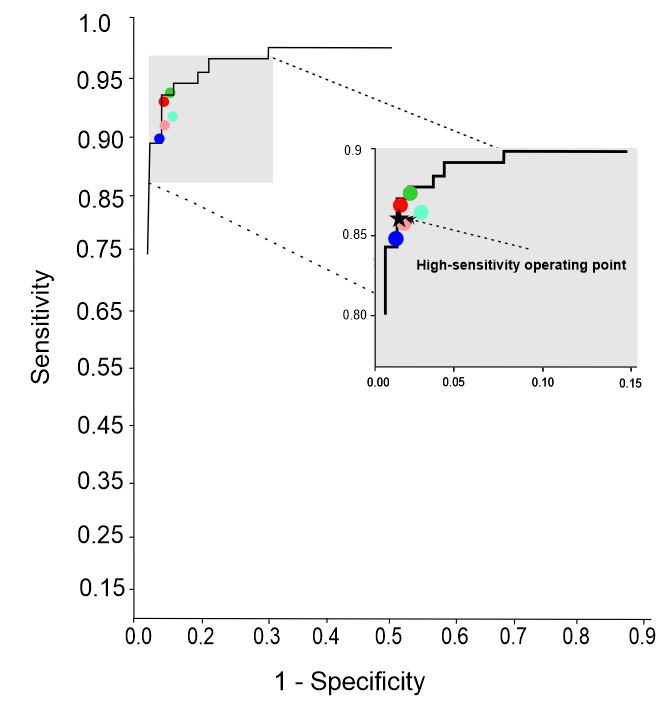}
    }
    \caption{ROC curves of the ensemble model~(black lines) for the detection of infection types on test set, color circles mean different folds, showing stable convergence} 
     \label{fig:rocs}
\end{figure}

Both DenseNets architectures show consistent performance owing to clearer image composition. DenseNet-161 outperforms not only DenseNet-201 but also all the other models. In particular, DenseNet-161 achieves precision, recall, and F1 scores of 0.952, 0.945, and 0.945, respectively, on balanced CXR images. VGG-19 also gives a comparable performance, giving precision, recall, and F1 scores of 0.943, 0.935, and 0.939, respectively. 
On imbalanced image sets, both DenseNet-161 and ResNet-18 perform consistently. Although VGG-19 and ResNet-18 show competitive results on the balanced dataset, the misclassification rate for normal and pneumonia samples are slightly elevated than DenseNet-161, which poses a risk for clinical diagnosis.
In contrast, DenseNet-161 is found to be resilient against class imbalanced. Hence, models like DenseNet-161, which can handle moderately imbalanced class scenarios, seem better suited for the clinical setting, where COVID-19 cases are rare compared to pneumonia or normal cases. The ROC curve of DenseNet-161 model in \cref{fig:rocs} shows consistent AUC scores across folds, indicating not only stable predictions but also better prediction than random guessing. 

\begin{figure*}
	\centering
	\includegraphics[width=0.85\linewidth,height=50mm]{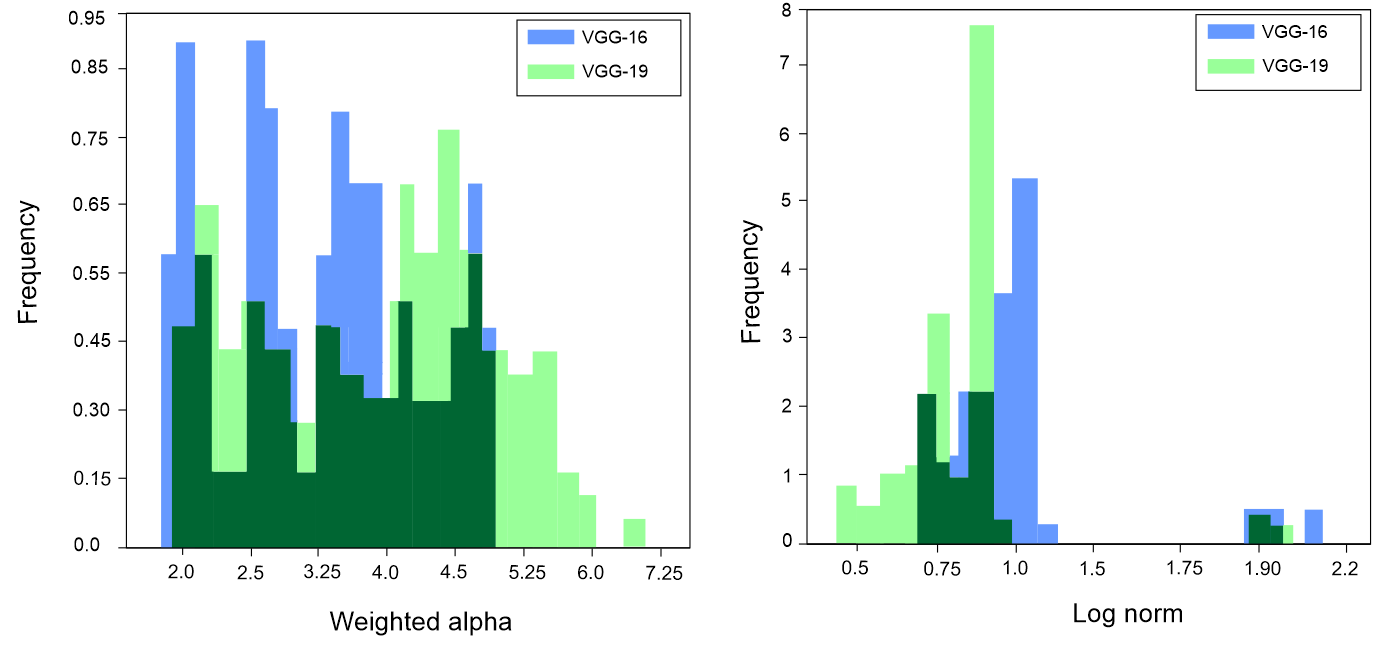}
	\caption{Choosing VGG model with WeightWatcher: a) w.r.t weighted alpha, b) w.r.t log-norm}
	\label{fig:weight_watch}
\end{figure*}

Nevertheless, lousy snapshot models can contaminate the overall predictive powers of the ensemble model. Hence, we employ WeightWatcher~\cite{martin2019traditional} in two levels: i)  level 1: we choose the top-5 snapshots to generate a full model, ii) level 2: we choose the top-3 models for the final ensemble model. In level 2, WeightWatcher is used to compare top models~(by excluding VGG-16, ResNet-34, and DenseNet-201) and choose the ones with the lowest log norm and highest weighted alpha~(refer to section 4 in supplementary for details), where a low~(weighted/average) log-norm signifies better generalization of network weights~\cite{martin2019traditional}. \Cref{fig:weight_watch} shows choosing the better model between VGG-16 and VGG-19 with WeightWatcher in terms of weighted alpha and log norm.

\subsection{Model ensemble}
\label{model_ensemble}
We perform the ensemble on following top-2 models: ResNet-18 and DenseNet-161. Besides, to ensure a variation of network architectures within the ensemble, VGG-19 is also included. 
As demonstrated in \cref{Table:ensemble_result}, the ensemble based on the SCPA method moderately outperforms the ensemble based on the PM method. The reason is that the PM approach appears to be easily influenced by outliers with high scores. To a great extent, the mean probabilities for each class affect the direction of outliers. For the SCPA-based ensemble, the combination of VGG-19 + DenseNet-161 outperforms other ensemble combination.  

\begin{figure}
   \centering
    \frame{
    \includegraphics[width=0.38\textwidth,height=45mm]{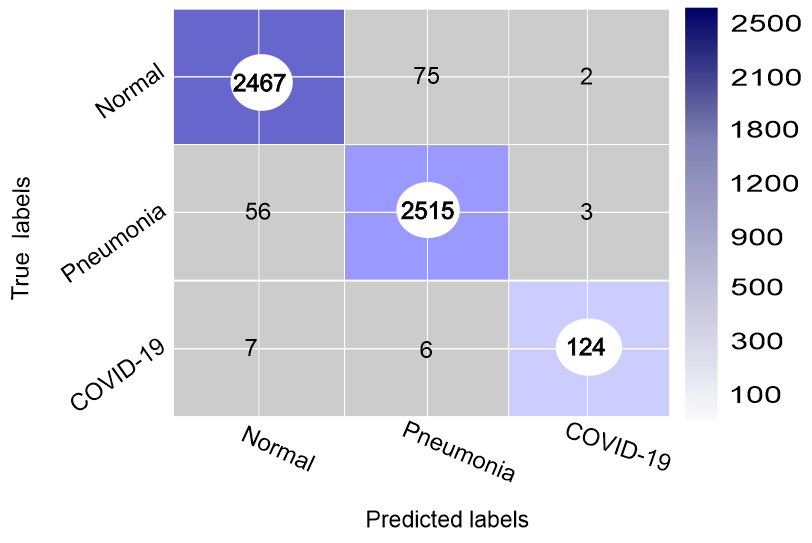}
    }
    \caption{confusion matrix of the ensemble model} 
     \label{fig:conf}
\end{figure}

The confusion matrix of the best ensemble's performance on balanced data is shown in \cref{fig:conf}. The results show that a majority of samples were classified correctly, with precision, recall, and F1 scores of 0.937, 0.926, and 0.931, respectively, using the PM ensemble method. For the SCPA-based ensemble, precision, recall, and F1 are even slightly higher, yielding 0.946, 0.943, and 0.945, respectively. Additionally, we report the class-specific measures in \cref{Table:class_specific_result} to give a better view in both the balanced and imbalanced scenario.

\begin{table*}
    \centering
    \caption{Classification results for ensemble methods on balanced dataset}
	\label{Table:ensemble_result}
	\begin{tabular}{p{4.3cm}p{1.6cm}p{1.1cm}p{0.8cm}p{0.1cm}p{1.6cm}p{1.1cm}p{0.8cm}}
		 &  \multicolumn{3}{c}{\textbf{Prediction maximization}} && \multicolumn{3}{c}{\textbf{Softmax posterior averaging}}  \\	
		\cmidrule{2-4}\cmidrule{6-8}
		\textbf{Architecture combination} &  \textbf{Precision} & \textbf{Recall}& \textbf{F1}&& \textbf{Precision} & \textbf{Recall}& \textbf{F1}\\
		\hline
		ResNet-18+DenseNet-161 & 0.915 & 0.924 & 0.928 && 0.925 & 0.94 & 0.933\\
		VGG-19+DenseNet-161 & \textbf{0.937} & \textbf{0.926} & \textbf{0.931} && \textbf{0.946} & \textbf{0.943} & \textbf{0.945}\\
		VGG-19+ResNet-18 & 0.917 & 0.923 & 0.912 &&  0.923 & 0.945 & 0.934\\
		DN-161+VGG-19+ResNet-18 & 0.926 & 0.901 & 0.901 && 0.924 & 0.937 & 0.935\\
		\hline 
	\end{tabular}
\end{table*}

\subsection{Quantitative analysis}
Since we primarily want to limit the number of missed COVID-19 instances, the achieved recall of 90.5\% is still an acceptable metric compared to 91\% by Wang et al.~\cite{wang2020covid}, which means that a certain fraction of all patients who test positive, will actually not have the disease. To determine how many of all infected persons would be diagnosed positive by the method, we calculate the positive predictive value~(PPV). Out of our test set with 129 COVID-19 patient samples, only 3 were misclassified as pneumonia and two as normal, which results in a PPV of 96.12\% for COVID-19 cases, which is still an acceptable metric compared to 98.9\% by Wang et al.~\cite{wang2020covid}. Besides, to provide a direct comparison, our approach slightly outperforming COVID-Net~\cite{wang2020covid} in terms of identifying normal and non-COVID19 cases. 

Further, to provide a one-to-one comparison with COVID-Net, we further apply similar techniques on the \emph{`COVIDx v1.0'} dataset. The evaluation result carried out on a test set of 92 COVID-19 cases, only 3 were misclassified as pneumonia, resulting in a PPV of 96.74\% for COVID-19 cases. This is, however, slightly lower than that of COVID-Net. However, we would still argue that it is an acceptable metric compared to 98.9\% by Wang et al.~\cite{wang2020covid}. It is important to note that PPV was reported for a low prevalence of COVID-19 in the cohorts. In a setting with high COVID-19 prevalence, the likelihood of false-positives is expected to reduce in favour of correct COVID-19 predictions. In our case, results are backed up by i) a larger test set, ii) better localization and explanation capability, which contributes to the reliability of our evaluation results~(refer to \cref{sub-sec:expl} for further details), given the fact that in healthcare predicting something with high confidence only is not enough, but requires trustworthiness. 

\begin{table*}
    \centering
	\caption{Classwise classification results of ensemble model on chest x-rays}
	\label{Table:class_specific_result}
	\begin{tabular}{p{3.2cm}p{1.7cm}p{1.1cm}p{1.3cm}p{0.1cm}p{1.7cm}p{1.2cm}p{1.0cm}}
		 & \multicolumn{3}{c}{\textbf{Balanced dataset}} && \multicolumn{3}{c}{\textbf{Imbalanced dataset}} \\	
		\cmidrule{2-4}\cmidrule{6-8}
		\textbf{Infection type} & \textbf{Precision} & \textbf{Recall}& \textbf{F1}&& \textbf{Precision} & \textbf{Recall}& \textbf{F1}\\	
		\hline
		Normal & 0.942 & 0.927 & 0.935 && 0.906 & 0.897 & 0.902\\
		Pneumonia & 0.916 & 0.928 & 0.922 && 0.864 & 0.853 & 0.858\\
		COVID-19 & 0.904 & 0.905 & 0.905 && 0.877 & 0.881 & 0.879\\
		\hline
	\end{tabular}
\end{table*}

\begin{figure*}
	\centering
    	\includegraphics[width=\textwidth,height=32mm]{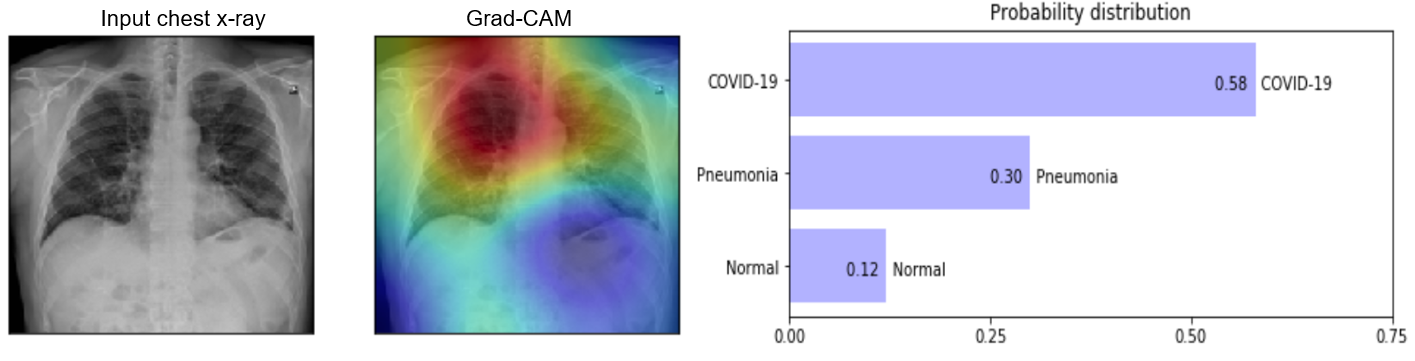}
    	\caption{The input chest x-ray classification, decision visualization with Grad-CAM and explanation}
    	\label{Fig:ggcam_viz}
    	\smallskip
    	\includegraphics[width=\textwidth,height=32mm]{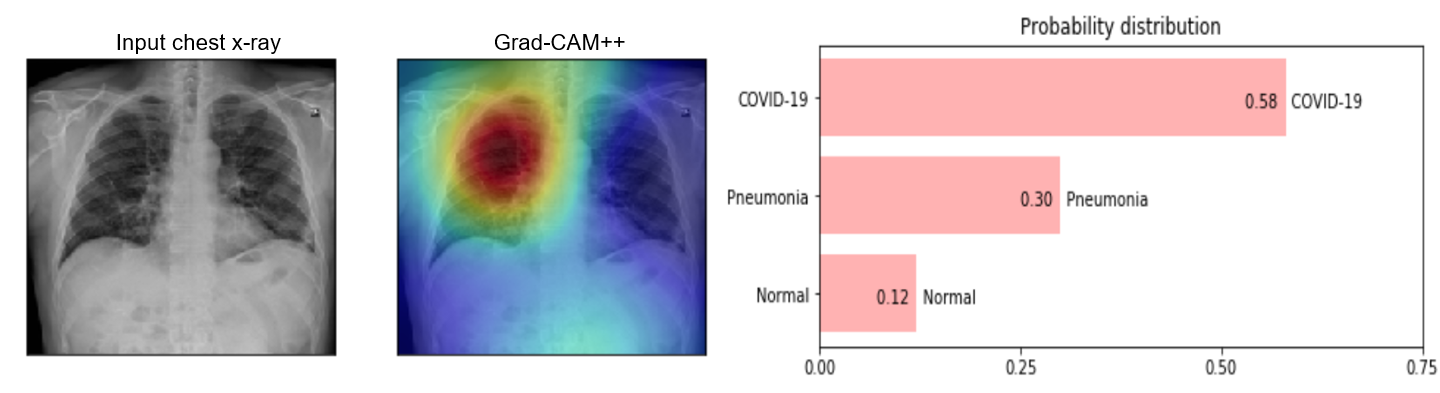}
    	\caption{The input chest x-ray classification, decision visualization with Grad-CAM++ and explanation}
    	\label{Fig:ggcam_plus_viz}
    	\smallskip
    	\includegraphics[width=\textwidth,height=32mm]{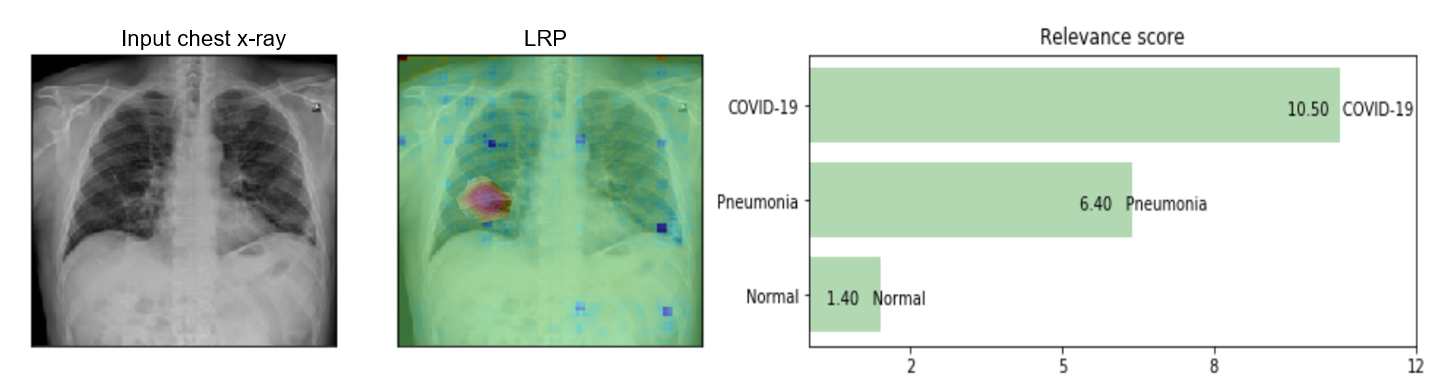}
    	\caption{The input chest x-ray classification, decision visualization with LRP and explanation}
    	\label{Fig:lrp_viz}
\end{figure*}

\subsection{COVID-19 predictions and explanations}
\label{sub-sec:expl}
Precise decisive feature localization is vital not only for the explanation but also for rapid confirmation of the reliability of outcomes, especially for potentially false-positive cases \cite{chattopadhay2018grad}. Attention map highlighting of critical regions on the chest advocate transparency and trustworthiness to clinicians and help them leverage their screening skills to make faster and yet more accurate diagnoses \cite{wang2020covid}. 
In general, the more accurate a model is, the more consistent the visualizations of Grad-CAM and Grad-CAM++ will be. Key features can then easily be identified based on where the activation maps are overlapping. The critical regions of some CXR images of COVID-19 cases are demonstrated in \cref{Fig:ggcam_viz}, \cref{Fig:ggcam_plus_viz}, and \cref{Fig:lrp_viz}, where class-discriminating areas within the lungs are localized.

As seen, HMs generated by Grad-CAM and Grad-CAM++ are fairly consistent and alike, but those with Grad-CAM++ are more accurately localized. The reason is that instead of certain parts, Grad-CAM++ highlights conjoined features more precisely.
On the other hand, although LRP highlights regions much more precisely, it fails to provide attention to critical regions. It turned out that Grad-CAM++ generates the most reliable HM's when compared to Grad-CAM and LRP. To provide more human-interpretable explanations, let's consider the following examples (based on ResNet-18): 

\begin{itemize}
    \item \textbf{Example 1}: the CXR image is classified to contain a confirmed COVID-19 case with a probability of 58\%, the true class is COVID-19, as shown in \cref{Fig:ggcam_viz}. 
    \item \textbf{Example 2}: the CXR image is classified to contain a confirmed COVID-19 case with a probability of 58\%, the true class is COVID-19, as shown in \cref{Fig:ggcam_plus_viz}. 
    \item \textbf{Example 3}: the CXR image is classified to contain COVID-19 case with a classification score of 10.5, the true class is COVID-19, as shown in \cref{Fig:lrp_viz}.
\end{itemize}

\subsection{Discussion and diagnosis recommendations}
Based on the above analyses, \emph{`DeepCOVIDExplainer'} disseminates the following recommendations: even if a specific approach does not perform well, an ensemble of several models still may outperform individual models. Since accurate diagnosis is a mandate, models trained on imbalanced training data may provide distorted or wrong predictions during inference time, due to possible overfitting during the training. In this case, even a high accuracy score can be achieved without predicting minor classes, hence might be uninformative. 

Thirdly, taking COVID-19 diagnosis context into account, the risk resulting from a pneumonia diagnosis is much lower than for a COVID-19 diagnosis. Hence, it is more reasonable to make a decision based on the maximum score among all single model predictions. Fourthly, due to the nature of neural networks, decision visualizations cannot be provided based on ensemble models, even though their usage contributes to decision fairness and reliability. For the decision visualization, therefore, it is recommended to pick the single best model as a basis and to employ Grad-CAM++ for providing the most reliable localization.


\section{Conclusion and Outlook}
\label{sec:co}
\noindent In this paper, we proposed \emph{`DeepCOVIDExplainer'} to leverage explainable COVID-19 prediction based on CXR images. Evaluation results show that our approach can identify COVID-19 with a PPV of 96.12\% and recall of 94.3\%, outperforming a recent approach. Further, as Curtis Langlotz\footnote{\url{https://www.nature.com/articles/d41586-019-03847-z}} stated ``AI won't replace radiologists, but radiologists who use AI will replace radiologists who don't''.  
However, we would argue that \emph{`DeepCOVIDExplainer'} is not to replace radiologists, instead of to be evaluated in a clinical setting and by no means a replacement for a human radiologist. In contrast, human judgement is indispensable when the life of patients is at stake. Further, we hope our findings will be a useful contribution to the fight against COVID-19 and towards an increasing acceptance and adoption of AI-assisted applications in the clinical practice. 

Lastly, we want to outline potential areas of improvements: 
first, since only a limited amount of CXR images for COVID-19 infection cases were at hand, it would be unfair to claim that we can rule out overfitting for our models. More unseen data from similar distributions is necessary for further evaluation to avoid possible out-of-distribution issues. 
Secondly, due to external conditions, we were yet not been able to verify the diagnoses and localization accuracies with the radiologists. 
Thirdly, accurate predictions do not only depend on single imaging modalities but could also build upon additional modalities like CT and other decisive factors such as e.g., patients demographic and symptomatic assessment report~\cite{7}. 
Nevertheless, we would argue that explaining predictions with plots and charts are useful for exploration and discovery. Explaining them to patients may be tedious and require more human-interpretable decision rules in natural language. In future, we intend to overcome these limitations by i) alleviating more data~(e.g., CT, phenotype, and history) and training a multimodal neural network, and ii) incorporating domain knowledge with symbolic reasoning to generate decision rules to make the diagnosis fairer.

\section*{Acknowledgments}
This work was supported by the German Ministry for Research and Education~(BMBF) as part of the SMITH consortium~(grant no. 01ZZ1803K). This work was conducted jointly by RWTH Aachen University and Fraunhofer FIT as part of the PHT and GoFAIR implementation network, which aims to develop a proof of concept information system to address current data reusability challenges of occurring in the context of so-called data integration centers that are being established as part of ongoing German Medical Informatics BMBF projects. 

\section*{Acronyms}
\noindent Acronyms and their full forms used in this paper are as follows:
\vspace{-3mm}

\begin{multicols}{2}
\begin{description}[leftmargin=0pt]
\scriptsize{
	\item[AUC] Area Under the receiver operating characteristic Curves
	 \item[CAM]  Class Activation Maps
	 \item[CXR]  Chest X-ray
	 \item[CNN]  Convolutional Neural Network
	 \item[CLRP]  Contrastive Layer-wise Relevance Propagation
	 \item[COVID-19]  Corona virus disease
	 \item[CT]  Computed Tomography
	 \item[CAC]  Cyclic Cosine Annealing
	 \item[DNN]  Deep Neural Networks
	 \item[DL]  Deep Learning 
	 \item[DenseNet]  Dense Convolutional Network
	 \item[DTD]  Deep Taylor Decomposition
	 \item[FCL]  Fully-Connected Layer
	 \item[FCN]  Fully Convolutional Neural Network
	 \item[FM]  Feature Maps
	 \item[GAG]  Globally Averaged Gradients 
	 \item[GAP]  Global Averaged Pooling 
	 \item[Grad-CAM]  Gradient-guided \\ class-activation maps
	 \item[HM]  Heat Maps
	 \item[HGE]  Histogram Equalization
	 \item[ICU]  Intensive Care Unit
	 \item[LR]  Learning Rate
	 \item[LRP]  Layer-wise relevance propagation
	 \item[ML]  Machine Learning  
	 \item[NE]  Number of Epochs
	 \item[PMF]  Perona-Malik Filter
	 \item[PM]  Prediction Maximization
	 \item[PPV]  Positive Predictive Value
	 \item[ROC]  Receiver Operating Characteristic
	 \item[ResNet]  Residual Network 
	 \item[RS]  Relevance Score
	 \item[RT-PCR] Reverse Transcriptase- \\polymerase Chain Reaction
	 \item[SARS-CoV-2]  Severe Acute Respiratory Syndrome Coronavirus 2
	 \item[SM]  Saliency Maps
	 \item[SGD]  Stochastic Gradient Descent
	 \item[SCPA]  Softmax Class Posterior Averaging
	 \item[SGLRP]  Softmax-gradient LRP
	 \item[TBF]  Tukey's biweight function
	 \item[TL]  Transfer Learning.
	 }
\end{description}
\end{multicols}

\bibliographystyle{ACM-Reference-Format}
\bibliography{Main.bib}


\begin{thebibliography}{33}


\ifx \showCODEN    \undefined \def \showCODEN     #1{\unskip}     \fi
\ifx \showDOI      \undefined \def \showDOI       #1{#1}\fi
\ifx \showISBNx    \undefined \def \showISBNx     #1{\unskip}     \fi
\ifx \showISBNxiii \undefined \def \showISBNxiii  #1{\unskip}     \fi
\ifx \showISSN     \undefined \def \showISSN      #1{\unskip}     \fi
\ifx \showLCCN     \undefined \def \showLCCN      #1{\unskip}     \fi
\ifx \shownote     \undefined \def \shownote      #1{#1}          \fi
\ifx \showarticletitle \undefined \def \showarticletitle #1{#1}   \fi
\ifx \showURL      \undefined \def \showURL       {\relax}        \fi
\providecommand\bibfield[2]{#2}
\providecommand\bibinfo[2]{#2}
\providecommand\natexlab[1]{#1}
\providecommand\showeprint[2][]{arXiv:#2}

\bibitem[\protect\citeauthoryear{Agarwal, Tiwari, and Lamba}{Agarwal
  et~al\mbox{.}}{2014}]%
        {90}
\bibfield{author}{\bibinfo{person}{Tarun~Kumar Agarwal},
  \bibinfo{person}{Mayank Tiwari}, {and} \bibinfo{person}{Subir~Singh Lamba}.}
  \bibinfo{year}{2014}\natexlab{}.
\newblock \showarticletitle{Modified histogram based contrast enhancement using
  homomorphic filtering for medical images}. In \bibinfo{booktitle}{\emph{2014
  IEEE International Advance Computing Conference (IACC)}}. IEEE,
  \bibinfo{pages}{964--968}.
\newblock


\bibitem[\protect\citeauthoryear{Ai, Yang, Hou, Zhan, and Xia}{Ai
  et~al\mbox{.}}{2020}]%
        {COVID3}
\bibfield{author}{\bibinfo{person}{Tao Ai}, \bibinfo{person}{Zhenlu Yang},
  \bibinfo{person}{Hongyan Hou}, \bibinfo{person}{Chenao Zhan}, {and}
  \bibinfo{person}{Liming Xia}.} \bibinfo{year}{2020}\natexlab{}.
\newblock \showarticletitle{Correlation of chest {CT} and {RT-PCR} testing in
  coronavirus disease 2019 ({COVID-19}) in {China}: a report of 1014 cases}.
\newblock \bibinfo{journal}{\emph{Radiology}} (\bibinfo{year}{2020}),
  \bibinfo{pages}{200642}.
\newblock


\bibitem[\protect\citeauthoryear{Baratloo, Hosseini, and El~Ashal}{Baratloo
  et~al\mbox{.}}{2015}]%
        {58}
\bibfield{author}{\bibinfo{person}{Alireza Baratloo}, \bibinfo{person}{Mostafa
  Hosseini}, {and} \bibinfo{person}{Gehad El~Ashal}.}
  \bibinfo{year}{2015}\natexlab{}.
\newblock \showarticletitle{Simple Definition and Calculation of Accuracy,
  Sensitivity and Specificity}.
\newblock \bibinfo{journal}{\emph{Emergency (Tehran, Iran)}}
  \bibinfo{volume}{3}, \bibinfo{number}{2} (\bibinfo{year}{2015}),
  \bibinfo{pages}{48--49}.
\newblock


\bibitem[\protect\citeauthoryear{Bullock, Pham, and Luengo-Oroz}{Bullock
  et~al\mbox{.}}{2020}]%
        {bullock2020mapping}
\bibfield{author}{\bibinfo{person}{Joseph Bullock},
  \bibinfo{person}{Katherine~Hoffmann Pham}, {and} \bibinfo{person}{Miguel
  Luengo-Oroz}.} \bibinfo{year}{2020}\natexlab{}.
\newblock \showarticletitle{Mapping the Landscape of Artificial Intelligence
  Applications against {COVID-19}}.
\newblock \bibinfo{journal}{\emph{arXiv:2003.11336}} (\bibinfo{year}{2020}).
\newblock


\bibitem[\protect\citeauthoryear{Candace and Daniel}{Candace and
  Daniel}{2020}]%
        {RT_PCR}
\bibfield{author}{\bibinfo{person}{Makeda~Moore Candace} {and}
  \bibinfo{person}{Daniel}.} \bibinfo{year}{2020}\natexlab{}.
\newblock \showarticletitle{COVID-19}.
\newblock  (\bibinfo{year}{2020}).
\newblock
\urldef\tempurl%
\url{https://radiopaedia.org/articles/covid-19-3?lang=us}
\showURL{%
\tempurl}
\newblock
\shownote{Online; accessed April-July-2020.}


\bibitem[\protect\citeauthoryear{Chattopadhay and Sarkar}{Chattopadhay and
  Sarkar}{2018}]%
        {chattopadhay2018grad}
\bibfield{author}{\bibinfo{person}{Aditya Chattopadhay} {and}
  \bibinfo{person}{Anirban Sarkar}.} \bibinfo{year}{2018}\natexlab{}.
\newblock \showarticletitle{Grad-{CAM}++: Generalized gradient-based visual
  explanations for convolutional networks}. In
  \bibinfo{booktitle}{\emph{Applications of Computer Vision(WACV)}}. IEEE,
  \bibinfo{pages}{839--847}.
\newblock


\bibitem[\protect\citeauthoryear{Cohen, Morrison, and Dao}{Cohen
  et~al\mbox{.}}{2020}]%
        {cohen2020covid}
\bibfield{author}{\bibinfo{person}{Joseph~Paul Cohen}, \bibinfo{person}{Paul
  Morrison}, {and} \bibinfo{person}{Lan Dao}.} \bibinfo{year}{2020}\natexlab{}.
\newblock \showarticletitle{{COVID-19} image data collection}.
\newblock \bibinfo{journal}{\emph{arXiv 2003.11597}} (\bibinfo{year}{2020}).
\newblock
\urldef\tempurl%
\url{https://github.com/ieee8023/covid-chestxray-dataset}
\showURL{%
\tempurl}


\bibitem[\protect\citeauthoryear{Fang, Zhang, Xie, Lin, Ying, Pang, and
  Ji}{Fang et~al\mbox{.}}{2020}]%
        {COVID2}
\bibfield{author}{\bibinfo{person}{Yicheng Fang}, \bibinfo{person}{Huangqi
  Zhang}, \bibinfo{person}{Jicheng Xie}, \bibinfo{person}{Minjie Lin},
  \bibinfo{person}{Lingjun Ying}, \bibinfo{person}{Peipei Pang}, {and}
  \bibinfo{person}{Wenbin Ji}.} \bibinfo{year}{2020}\natexlab{}.
\newblock \showarticletitle{Sensitivity of chest {CT} for {COVID-19}:
  comparison to {RT-PCR}}.
\newblock \bibinfo{journal}{\emph{Radiology}} (\bibinfo{year}{2020}),
  \bibinfo{pages}{200432}.
\newblock


\bibitem[\protect\citeauthoryear{Ghoshal and Tucker}{Ghoshal and
  Tucker}{2020}]%
        {ghoshal2020estimating}
\bibfield{author}{\bibinfo{person}{Biraja Ghoshal} {and} \bibinfo{person}{Allan
  Tucker}.} \bibinfo{year}{2020}\natexlab{}.
\newblock \showarticletitle{Estimating Uncertainty and Interpretability in Deep
  Learning for Coronavirus ({COVID-19}) Detection}.
\newblock \bibinfo{journal}{\emph{arXiv:2003.10769}} (\bibinfo{year}{2020}).
\newblock


\bibitem[\protect\citeauthoryear{Gozes, Frid-Adar, Greenspan, and Siegel}{Gozes
  et~al\mbox{.}}{2020}]%
        {gozes2020rapid}
\bibfield{author}{\bibinfo{person}{Ophir Gozes}, \bibinfo{person}{Maayan
  Frid-Adar}, \bibinfo{person}{Hayit Greenspan}, {and} \bibinfo{person}{Eliot
  Siegel}.} \bibinfo{year}{2020}\natexlab{}.
\newblock \showarticletitle{Rapid {AI} development cycle for the coronavirus
  pandemic: Initial results for automated detection and patient monitoring
  using deep learning {CT} image analysis}.
\newblock \bibinfo{journal}{\emph{arXiv:2003.05037}} (\bibinfo{year}{2020}).
\newblock


\bibitem[\protect\citeauthoryear{He, Zhang, Ren, and Sun}{He
  et~al\mbox{.}}{2016}]%
        {106}
\bibfield{author}{\bibinfo{person}{Kaiming He}, \bibinfo{person}{Xiangyu
  Zhang}, \bibinfo{person}{Shaoqing Ren}, {and} \bibinfo{person}{Jian Sun}.}
  \bibinfo{year}{2016}\natexlab{}.
\newblock \showarticletitle{Deep residual learning for image recognition}. In
  \bibinfo{booktitle}{\emph{Proc. of the IEEE CVPR}}.
  \bibinfo{pages}{770--778}.
\newblock


\bibitem[\protect\citeauthoryear{Huang, Wang, Li, Ren, and Gu}{Huang
  et~al\mbox{.}}{2020}]%
        {huang2020clinical}
\bibfield{author}{\bibinfo{person}{Chaolin Huang}, \bibinfo{person}{Yeming
  Wang}, \bibinfo{person}{Xingwang Li}, \bibinfo{person}{Lili Ren}, {and}
  \bibinfo{person}{Xiaoying Gu}.} \bibinfo{year}{2020}\natexlab{}.
\newblock \showarticletitle{Clinical features of patients infected with novel
  coronavirus in {Wuhan, China}}.
\newblock \bibinfo{journal}{\emph{The Lancet}} \bibinfo{volume}{395},
  \bibinfo{number}{10223} (\bibinfo{year}{2020}), \bibinfo{pages}{497--506}.
\newblock


\bibitem[\protect\citeauthoryear{Huang, Li, Pleiss, and Weinberger}{Huang
  et~al\mbox{.}}{2017a}]%
        {huang2017snapshot}
\bibfield{author}{\bibinfo{person}{Gao Huang}, \bibinfo{person}{Yixuan Li},
  \bibinfo{person}{Geoff Pleiss}, {and} \bibinfo{person}{Kilian~Q Weinberger}.}
  \bibinfo{year}{2017}\natexlab{a}.
\newblock \showarticletitle{Snapshot Ensembles: Train 1, get m for free}.
\newblock \bibinfo{journal}{\emph{arXiv:1704.00109}} (\bibinfo{year}{2017}).
\newblock


\bibitem[\protect\citeauthoryear{Huang, Liu, and Weinberger}{Huang
  et~al\mbox{.}}{2017b}]%
        {109}
\bibfield{author}{\bibinfo{person}{Gao Huang}, \bibinfo{person}{Zhuang Liu},
  {and} \bibinfo{person}{Kilian~Q Weinberger}.}
  \bibinfo{year}{2017}\natexlab{b}.
\newblock \showarticletitle{Densely connected convolutional networks}. In
  \bibinfo{booktitle}{\emph{proc of the IEEE CVPR}}.
  \bibinfo{pages}{4700--4708}.
\newblock


\bibitem[\protect\citeauthoryear{Iwana, Kuroki, and Uchida}{Iwana
  et~al\mbox{.}}{2019}]%
        {LRP2}
\bibfield{author}{\bibinfo{person}{Brian~Kenji Iwana}, \bibinfo{person}{Ryohei
  Kuroki}, {and} \bibinfo{person}{Seiichi Uchida}.}
  \bibinfo{year}{2019}\natexlab{}.
\newblock \showarticletitle{Explaining Convolutional Neural Networks using
  Softmax Gradient Layer-wise Relevance Propagation}.
\newblock \bibinfo{journal}{\emph{arXiv:1908.04351}} (\bibinfo{year}{2019}).
\newblock


\bibitem[\protect\citeauthoryear{Kamalaveni, Rajalakshmi, and
  Narayanankutty}{Kamalaveni et~al\mbox{.}}{2015}]%
        {95}
\bibfield{author}{\bibinfo{person}{V Kamalaveni}, \bibinfo{person}{R~Anitha
  Rajalakshmi}, {and} \bibinfo{person}{KA Narayanankutty}.}
  \bibinfo{year}{2015}\natexlab{}.
\newblock \showarticletitle{Image denoising using variations of Perona-Malik
  model with different edge stopping functions}.
\newblock \bibinfo{journal}{\emph{Procedia Computer Science}}
  \bibinfo{volume}{58} (\bibinfo{year}{2015}), \bibinfo{pages}{673--682}.
\newblock


\bibitem[\protect\citeauthoryear{Karim, Rahman, Jares, Decker, and Beyan}{Karim
  et~al\mbox{.}}{2019}]%
        {karim2019snapshot}
\bibfield{author}{\bibinfo{person}{Md.~Rezaul Karim}, \bibinfo{person}{Ashiqur
  Rahman}, \bibinfo{person}{Jo{\~a}o~Bosco Jares}, \bibinfo{person}{Stefan
  Decker}, {and} \bibinfo{person}{Oya Beyan}.} \bibinfo{year}{2019}\natexlab{}.
\newblock \showarticletitle{{A Snapshot Neural Ensemble Method for Cancer-type
  Prediction Based on Copy Number Variations}}.
\newblock \bibinfo{journal}{\emph{Neural Computing and Applications}}
  (\bibinfo{year}{2019}), \bibinfo{pages}{1--19}.
\newblock


\bibitem[\protect\citeauthoryear{Kim, Koo, Choi, and Kim}{Kim
  et~al\mbox{.}}{2020}]%
        {kim2020extending}
\bibfield{author}{\bibinfo{person}{Bum~Jun Kim}, \bibinfo{person}{Gyogwon Koo},
  \bibinfo{person}{Hyeyeon Choi}, {and} \bibinfo{person}{Sang~Woo Kim}.}
  \bibinfo{year}{2020}\natexlab{}.
\newblock \showarticletitle{Extending Class Activation Mapping Using Gaussian
  Receptive Field}.
\newblock \bibinfo{journal}{\emph{arXiv:2001.05153}} (\bibinfo{year}{2020}).
\newblock


\bibitem[\protect\citeauthoryear{Loshchilov and Hutter}{Loshchilov and
  Hutter}{2016}]%
        {loshchilov2016sgdr}
\bibfield{author}{\bibinfo{person}{Ilya Loshchilov} {and}
  \bibinfo{person}{Frank Hutter}.} \bibinfo{year}{2016}\natexlab{}.
\newblock \showarticletitle{{SGDR: Stochastic Gradient Descent with Warm
  Restarts}}.
\newblock \bibinfo{journal}{\emph{arXiv:1608.03983}} (\bibinfo{year}{2016}).
\newblock


\bibitem[\protect\citeauthoryear{Martin and Mahoney}{Martin and
  Mahoney}{2019}]%
        {martin2019traditional}
\bibfield{author}{\bibinfo{person}{Charles~H Martin} {and}
  \bibinfo{person}{Michael~W Mahoney}.} \bibinfo{year}{2019}\natexlab{}.
\newblock \showarticletitle{Traditional and heavy-tailed self regularization in
  neural network models}.
\newblock \bibinfo{journal}{\emph{arXiv:1901.08276}} (\bibinfo{year}{2019}).
\newblock


\bibitem[\protect\citeauthoryear{Montavon, Lapuschkin, and M{\"u}ller}{Montavon
  et~al\mbox{.}}{2017}]%
        {DTD}
\bibfield{author}{\bibinfo{person}{Gr{\'e}goire Montavon},
  \bibinfo{person}{Sebastian Lapuschkin}, {and} \bibinfo{person}{Klaus-Robert
  M{\"u}ller}.} \bibinfo{year}{2017}\natexlab{}.
\newblock \showarticletitle{Explaining nonlinear classification decisions with
  deep taylor decomposition}.
\newblock \bibinfo{journal}{\emph{Pattern Recognition}}  \bibinfo{volume}{65}
  (\bibinfo{year}{2017}), \bibinfo{pages}{211--222}.
\newblock


\bibitem[\protect\citeauthoryear{Narin, Kaya, and Pamuk}{Narin
  et~al\mbox{.}}{2020}]%
        {narin2020automatic}
\bibfield{author}{\bibinfo{person}{Ali Narin}, \bibinfo{person}{Ceren Kaya},
  {and} \bibinfo{person}{Ziynet Pamuk}.} \bibinfo{year}{2020}\natexlab{}.
\newblock \showarticletitle{Automatic Detection of Coronavirus Disease using
  {X-ray} Images and Deep Convolutional Neural Networks}.
\newblock \bibinfo{journal}{\emph{arXiv:2003.10849}} (\bibinfo{year}{2020}).
\newblock


\bibitem[\protect\citeauthoryear{Ng, Lee, Yang, and Khong}{Ng
  et~al\mbox{.}}{2020}]%
        {ng2020imaging}
\bibfield{author}{\bibinfo{person}{Ming-Yen Ng}, \bibinfo{person}{Elaine~Y
  Lee}, \bibinfo{person}{Jin Yang}, {and} \bibinfo{person}{Pek-Lan Khong}.}
  \bibinfo{year}{2020}\natexlab{}.
\newblock \showarticletitle{Imaging profile of {COVID-19} infection: radiologic
  findings and literature review}.
\newblock \bibinfo{journal}{\emph{Cardiothoracic Imaging}} \bibinfo{volume}{2},
  \bibinfo{number}{1} (\bibinfo{year}{2020}).
\newblock


\bibitem[\protect\citeauthoryear{Ozturk, Talo, Yildirim, Baloglu, Yildirim, and
  Acharya}{Ozturk et~al\mbox{.}}{2020}]%
        {ozturk2020automated}
\bibfield{author}{\bibinfo{person}{Tulin Ozturk}, \bibinfo{person}{Muhammed
  Talo}, \bibinfo{person}{Eylul~Azra Yildirim}, \bibinfo{person}{Ulas~Baran
  Baloglu}, \bibinfo{person}{Ozal Yildirim}, {and} \bibinfo{person}{U~Rajendra
  Acharya}.} \bibinfo{year}{2020}\natexlab{}.
\newblock \showarticletitle{Automated detection of COVID-19 cases using deep
  neural networks with X-ray images}.
\newblock \bibinfo{journal}{\emph{Computers in Biology and Medicine}}
  (\bibinfo{year}{2020}), \bibinfo{pages}{103792}.
\newblock


\bibitem[\protect\citeauthoryear{Pathak, Dahiwale, and Padole}{Pathak
  et~al\mbox{.}}{2015}]%
        {86}
\bibfield{author}{\bibinfo{person}{Sampada~S Pathak}, \bibinfo{person}{Prashant
  Dahiwale}, {and} \bibinfo{person}{Ganesh Padole}.}
  \bibinfo{year}{2015}\natexlab{}.
\newblock \showarticletitle{A combined effect of local and global method for
  contrast image enhancement}. In \bibinfo{booktitle}{\emph{International
  Conference on Engineering \& Technology}}. IEEE, \bibinfo{pages}{1--5}.
\newblock


\bibitem[\protect\citeauthoryear{Perona and Malik}{Perona and Malik}{1990}]%
        {96}
\bibfield{author}{\bibinfo{person}{Pietro Perona} {and}
  \bibinfo{person}{Jitendra Malik}.} \bibinfo{year}{1990}\natexlab{}.
\newblock \showarticletitle{Scale-space and edge detection using anisotropic
  diffusion}.
\newblock \bibinfo{journal}{\emph{IEEE Tras. on pattern analysis and machine
  intelligence}} \bibinfo{volume}{12}, \bibinfo{number}{7}
  (\bibinfo{year}{1990}), \bibinfo{pages}{629--639}.
\newblock


\bibitem[\protect\citeauthoryear{Selvaraju, Cogswell, Das, Vedantam, Parikh,
  and Batra}{Selvaraju et~al\mbox{.}}{2017}]%
        {114}
\bibfield{author}{\bibinfo{person}{Ramprasaath~R Selvaraju},
  \bibinfo{person}{Michael Cogswell}, \bibinfo{person}{Abhishek Das},
  \bibinfo{person}{Ramakrishna Vedantam}, \bibinfo{person}{Devi Parikh}, {and}
  \bibinfo{person}{Dhruv Batra}.} \bibinfo{year}{2017}\natexlab{}.
\newblock \showarticletitle{{Grad-CAM}: Visual explanations from deep networks
  via gradient-based localization}. In \bibinfo{booktitle}{\emph{Proceedings of
  the IEEE ICCV}}. \bibinfo{pages}{618--626}.
\newblock


\bibitem[\protect\citeauthoryear{Simonyan and Zisserman}{Simonyan and
  Zisserman}{2014}]%
        {108}
\bibfield{author}{\bibinfo{person}{Karen Simonyan} {and}
  \bibinfo{person}{Andrew Zisserman}.} \bibinfo{year}{2014}\natexlab{}.
\newblock \showarticletitle{Very deep convolutional networks for large-scale
  image recognition}.
\newblock \bibinfo{journal}{\emph{arXiv:1409.1556}} (\bibinfo{year}{2014}).
\newblock


\bibitem[\protect\citeauthoryear{Tabik, G{\'o}mez-R{\'\i}os,
  Mart{\'\i}n-Rodr{\'\i}guez, Sevillano-Garc{\'\i}a, Rey-Area, Charte, Guirado,
  Su{\'a}rez, Luengo, Valero-Gonz{\'a}lez, et~al\mbox{.}}{Tabik
  et~al\mbox{.}}{2020}]%
        {tabik2020covidgr}
\bibfield{author}{\bibinfo{person}{S Tabik}, \bibinfo{person}{A
  G{\'o}mez-R{\'\i}os}, \bibinfo{person}{JL Mart{\'\i}n-Rodr{\'\i}guez},
  \bibinfo{person}{I Sevillano-Garc{\'\i}a}, \bibinfo{person}{M Rey-Area},
  \bibinfo{person}{D Charte}, \bibinfo{person}{E Guirado}, \bibinfo{person}{JL
  Su{\'a}rez}, \bibinfo{person}{J Luengo}, \bibinfo{person}{MA
  Valero-Gonz{\'a}lez}, {et~al\mbox{.}}} \bibinfo{year}{2020}\natexlab{}.
\newblock \showarticletitle{COVIDGR dataset and COVID-SDNet methodology for
  predicting COVID-19 based on Chest X-Ray images}.
\newblock \bibinfo{journal}{\emph{arXiv preprint arXiv:2006.01409}}
  (\bibinfo{year}{2020}).
\newblock


\bibitem[\protect\citeauthoryear{Tiulpin, Thevenot, Rahtu, Lehenkari, and
  Saarakkala}{Tiulpin et~al\mbox{.}}{2018}]%
        {7}
\bibfield{author}{\bibinfo{person}{Aleksei Tiulpin},
  \bibinfo{person}{J{\'e}r{\^o}me Thevenot}, \bibinfo{person}{Esa Rahtu},
  \bibinfo{person}{Petri Lehenkari}, {and} \bibinfo{person}{Simo Saarakkala}.}
  \bibinfo{year}{2018}\natexlab{}.
\newblock \showarticletitle{Automatic Knee Osteoarthritis Diagnosis from Plain
  Radiographs: A Deep Learning-based Approach}.
\newblock \bibinfo{journal}{\emph{Scientific reports}} \bibinfo{volume}{8},
  \bibinfo{number}{1} (\bibinfo{year}{2018}), \bibinfo{pages}{1727}.
\newblock


\bibitem[\protect\citeauthoryear{Wang and Wong}{Wang and Wong}{2020}]%
        {wang2020covid}
\bibfield{author}{\bibinfo{person}{Linda Wang} {and} \bibinfo{person}{Alexander
  Wong}.} \bibinfo{year}{2020}\natexlab{}.
\newblock \showarticletitle{COVID-Net: A Tailored Deep Convolutional Neural
  Network Design for Detection of {COVID-19} Cases from Chest {X-ray} Images}.
\newblock \bibinfo{journal}{\emph{arXiv:2003.09871}} (\bibinfo{year}{2020}).
\newblock


\bibitem[\protect\citeauthoryear{Xie, Girshick, and He}{Xie
  et~al\mbox{.}}{2017}]%
        {107}
\bibfield{author}{\bibinfo{person}{Saining Xie}, \bibinfo{person}{Ross
  Girshick}, {and} \bibinfo{person}{Kaiming He}.}
  \bibinfo{year}{2017}\natexlab{}.
\newblock \showarticletitle{Aggregated residual transformations for deep neural
  networks}. In \bibinfo{booktitle}{\emph{proc. of IEEE conference on computer
  vision and pattern recognition}}. \bibinfo{pages}{1492--1500}.
\newblock


\bibitem[\protect\citeauthoryear{Yee}{Yee}{2020}]%
        {COVID1}
\bibfield{author}{\bibinfo{person}{Kate~Madden Yee}.}
  \bibinfo{year}{2020}\natexlab{}.
\newblock \showarticletitle{X-ray may be missing {COVID} cases found with
  {CT}}.
\newblock \bibinfo{journal}{\emph{Korean Journal of Radiology}}
  (\bibinfo{year}{2020}), \bibinfo{pages}{1--7}.
\newblock


\end{thebibliography}

\end{document}